\documentclass[12pt]{article}
\usepackage{epsf}
\usepackage[dvips]{graphicx}

\begin{document}
\pagestyle{empty}

\begin{center}
{\bfseries STUDY OF MULTIPARTICLE PRODUCTION BY GLUON DOMINANCE
MODEL (Part II)} \footnote {Talk given at XVII International
Baldin - Seminar "Relativistic Nuclear Physics and Quantum
Chromodynamics". JINR. September 27 - October 2, 2004, Dubna,
Russia.}
 \vskip 5mm

P.F.~Ermolov$^1$, E.S.~Kokoulina$^{2,3 \dag}$, E.A.~Kuraev$^3$
A.V.~Kutov$^3$,

V.A.~Nikitin$^3$, A.A.~Pankov$^2$, I.A.~Roufanov$^{3}$,
N.K.~Zhidkov$^3$ \vskip 5mm
 {\small (1) {\it SINP MSU},
(2) {\it GSTU, Belarus }, (3) {\it JINR }
\\
$\dag$ {\it E-mail: Elena.Kokoulina@sunse.jinr.ru }}
\end{center}

\vskip 5mm

\begin{center}
\begin{minipage}{150mm}
\centerline{\bf Abstract}

The gluon dominance model presents a description of multiparticle
production in proton-proton collisions and proton-antiproton
annihilation. The collective behavior of secondary particles in
$pp$-interactions at 70 GeV/c and higher is studied in the project
{\bf "Thermalization"}. The obtained neutral and charged
multiplicity distribution parameters explain some RHIC-data. The
gluon dominance model is  modified by the inclusion of
intermediate quark topology for the multiplicity distribution
description in the pure $p\bar p$-annihilation at few tens GeV/c
and explains behavior of the second correlative moment. This
article proposes a mechanism of the soft photon production as a
sign of hadronization. Excess of soft photons allows one to
estimate the emission region size.
\end{minipage}
\end{center}

\vskip 5mm

\section{Introduction}

A new model of investigating multiparticle production (MP) at high
energy is proposed. It is based on multiplicity distribution (MD)
description of different interactions on basis of QCD and a
phenomenological hadronization scheme. It is shown that the
proposed model agrees with experimental data in a wide energy
region and, perhaps, can be used for analysis of jet quenching and
other phenomena at RHIC \cite{WhPa}.

Application of this model approach to pp-interaction (for the
beginning see \cite{BAL}) is given in Section 2. The additional
investigations of MD in the $p \bar p $ annihilation channel at a
few tens GeV/c are carried out in Section 3. The emission region
size for soft photons and the possible mechanisms of their
formation are discussed in Section 4. The main results of these
studies are given in Section 5.

\section{MD in $pp$-interactions (continuation)}

MD of charged particles in proton interactions by means of the
gluon dominance model were studied in \cite{BAL}. It is
interesting to get MD for neutral mesons. For this purpose we take
experimental mean multiplicity of $\pi ^0$ in pp-interactions at
69 GeV/c ($\sqrt s\simeq $ 11.6~ GeV). It was be found $2.57\pm
0.13$ \cite{MUR}. So the mean multiplicity in this process is
calculated as the product of the mean number of evaporated active
gluons ($\overline m=2.48$) and hadron parameter $\overline n^h$.
We can determine the hadronization parameter for neutral mesons:
$\overline n^h_0$=$1.036\pm 0.041$ \cite{MGD}. We expect
approximate equality of probabilities of different hadron
production at the second (hadronization) stage. MD for neutral
mesons have a form as for charged particles \cite{BAL}:
\begin{equation}
\label{38} P_n(s)=\sum\limits_{m=0}^{ME}\frac{e^{-\overline m}
\overline m^m}{m!} C^{n-2}_{mN}\left(\frac{\overline n^h}
{N}\right)^{n-2}\left(1-\frac{\overline n^h}
{N}\right)^{mN-(n-2)},
\end{equation}
and can be easily obtained if they are normalized to mean
multiplicity $\pi ^0$'s (Fig. 1). From this distribution we see
that the maximal possible number of $\pi ^0$ from TSTM \cite{MGD}
is 16. MD for the total multiplicity are shown in Fig. 2. The
maximal total number of particles in this case is equal to 42.

The dependence of the mean multiplicity of neutral mesons
$\overline n_0$ versus the number of charged particles $n_{ch}$
can be determined  by means of MD $P_{n_{tot}}(s)$:

\begin{equation}
\label{40} \overline n_0(n_{ch},s)=\frac {\sum \limits
_{n_{tot}=n_1}^{n_2} P_{n_{tot}}(s) \cdot (n_{tot}-n_{ch})}{\sum
\limits _{n_{tot}=n_1}^{n_2} P_{n_{tot}}(s)},
\end{equation}
where $n_1$ and $n_2$ are lower and top boundaries for the total
multiplicity at the given number of charged particles $n_{ch}$.
The MD of charged and neutral secondaries obtained by TSTM give
the maximal number for charged $n_{ch}=26$, neutral $n_{0}=16$ and
total $n_{tot}=42$. That is why we have the following limits for
$n_1$ and $n_2$: $n_1 \geq n_{ch}$, $n_2 \leq 16+ n_{ch}$. These
restrictions result in great disagreement with experimental data
\cite{MUR} at small multiplicities. It was shown in \cite{MGD}.

A significant improvement will be reached if we decrease the top
limit at low multiplicities ($n_{ch}\leq 10$) to $n_2=2 n_{ch}$.
This corresponds to the case when the maximal number of neutrals
is equal to the number of charged particles, and a double excess
of neutral mesons over positive (negative) pions is possible. Fig.
3 shows that multiplicity of neutrals versus $n_{ch}$ when $n_2$
is taken equal to $2 n_{ch}$ at small $n_{ch}$ and $n_2=16+n_{ch}$
at $n_{ch}>10$. This restriction in (\ref{40}) indicates that
AntiCentauro events (a large number of neutrals and very few
charged particles) must be absent. Centauro events (a large number
of charged particles and practically no accompanying neutrals) may
be realized only in the region of high multiplicity.

It is assumed \cite{KUR} that at the second stage different kinds
of quark pairs from the gluon (maximal possible number is equal to
$N_{tot}$) occur with equal probabilities. We will try to consider
the formation of neutral and charged mesons as an example of the
above assumption. The $u \overline u$ and $d \overline d$ quark
pairs may appear at sufficient energy. At the end of hadronization
the formation of two charged mesons (the law of charge
conservation of quarks) may take place. Production of an
additional neutral particle is not necessary while formation of a
neutral meson. So we can claim that the number of charged hadrons
will be larger than the number of the neutral ones, or the
probability of the charged hadron production is higher than of the
neutral ones. We can estimate these probabilities in GDM.

MD of $\pi ^0$ from one gluon at the second stage may be described
by the binomial distribution $P_{n_0}=
C_{n_t}^{n_0}p_0^{n_0}p_c^{n_t-n_0}$. Here $n_t$ is the total
number of hadrons formed from gluon, $n_0$ - the number of neutral
mesons among these secondaries (the number of charged hadrons
$n_c=n_t-n_0$), $p_c (p_0)$ -the probability of production of
charged pair (one $\pi^0$). The normalized condition is
$p_0+p_c=1$. From TSTM we have obtained $\overline n_{ch}=1.63$
and $\overline n_{0}=1.036$. The mean multiplicities for binomial
distributions will be equal to: $\overline n_{ch}=p_c \overline
n_t$, $\overline n_{0} =p_0 \overline n_t$. The probability of the
charge particle production is higher than of the neutral mesons
($\overline n_{ch} > \overline n_{0}$). The ratio of these values
is $p_c/p_0\sim 1.46$.

The mean multiplicity of newly born hadrons (charged or neutral)
in proton interactions in GDM is equal to the product of the mean
multiplicity of gluons obtained at the first stage and the mean
multiplicity of hadrons ($\overline n^h_{ch}$ or $\overline
n^h_{0}$) produced from one gluon at the second stage. In the case
of binomial distribution $\overline n_{ch}=\overline n_t\cdot
p_c$, $\overline n_{0}=\overline n_t \cdot p_0$. Taking into
account two leading protons, the mean multiplicity is $ \overline
n_{ch}(s)=2+\overline m_g (s)\cdot \overline n^h_{ch}$ for charged
particles in pp-interactions. The mean multiplicity of neutral
mesons in this process is $ \overline n_{0}(s)=\overline m_g(s)
\cdot \overline n^h_{0}.$ The ratio of the mean charged pairs to
the neutral mesons in proton interactions is
\begin{equation}
\label{43} \frac {\overline n_{ch}(s)/2}{\overline n_{0}}= \frac
{1}{\overline m(s)\cdot\overline n_0^h}+\frac {1}{2}\cdot\frac
{\overline n_{ch}^h}{\overline n_0^h}.
\end{equation}
At 69 GeV/c this ratio (\ref{43}) is equal to $1.19 \pm .25$. At
the higher energy the mean number of active gluons $\overline m$
increases  and becomes much more than 3. In this case (\ref{43})
it will be around the ratio of $\overline n_{ch}^h/ 2\overline
n_0^h$. The experimental data have shown $1.6$ for Au-Au
peripheral interactions ($80-92(\%)$ centrality class) at 200 GeV
and for pp interactions at 53 GeV \cite{PHE}. We can compare these
results with GDM at higher energies.

The application of GDM to describe MD in the energy region (102,
205, 300, 405 and 800 GeV/c) \cite{PrD} in both schemes (TSMB and
TSTM) \cite{BAL} leads to good results (Fig. 4-8). Parameters of
TSTM in this domain are given in Table 2.

\begin{center}
Table 1. Parameters of TSTM.
\end{center}
\renewcommand{\tablename}{Table}
\begin{center}
\begin{tabular}{||c||c|c|c|c|c|c||c||}
\hline \hline
$\sqrt s$ GeV & $\overline m$ & $M_g$ & $N$&$\overline n^h$&$\Omega $&$\chi^2/$ndf\\
\hline \hline
102  & $2.75\pm 0.08$ & 8  & $3.13\pm 0.56$ & $1.64 \pm 0.04$ & $1.92\pm 0.08$ & 2.2/5\\
\hline
205  & $2.82 \pm 0.20$ & 8 & $4.50 \pm 0.10$& $2.02 \pm 0.12$& $2.00 \pm 0.07$  &2.0/8  \\
\hline
300& $2.94 \pm 0.34$ & 10  &$4.07 \pm 0.86$ & $2.22 \pm 0.23$& $1.97 \pm 0.05$  & 9.8/9 \\
\hline
405& $2.70 \pm 0.30$& 9& $4.60 \pm 0.24$ & $2.66 \pm 0.22$ & $1.98 \pm 0.07$ &16.4/12 \\
\hline
800  &  $3.41 \pm 2.55$& 10&$ 20.30 \pm 10.40 $& $2.41 \pm 1.69$ & $2.01 \pm 0.08$&10.8/12\\
\hline \hline
\end{tabular}
\end{center}

We see that the number of active gluons and their mean
multiplicity increase, parameters of hadronization $N$ and
$\overline n^h_{ch}$ vary very slowly. At these energies the
charged hadron/pion ratio (\ref{43})  grows up to 1.6. The
parameter of hadronization $\overline n_{ch}^h$ has a trend to
increase weakly but $\overline n_{0}^h$ does not almost change.
This behavior may be related with the production of other charged
particles (not only pions): protons, antiprotons, kaons and so on.
We consider that parameter $\overline n^h_{ch}$ goes to the limit
value (like saturation).

On the other side a small growth $\overline n^h_g$ in proton
interactions also points at a possible change mechanism of
hadronization of gluons in comparison with the transition gluons
to hadrons in $e^+e^-$ annihilation. It is considered that in the
last case partons transform to hadrons by the fragmentation
mechanism at the absence of the thermal medium. Our MD analysis
gives $\overline n_g^h \sim 1$ for this fragmentation \cite{ALU}.
The recombination is specific for the hadron and nucleus
processes. In this situation a lot of quark pairs from gluons
appear almost simultaneously and recombine to various hadrons
\cite{HWA}. The value $\overline n_g^h$ becomes bigger $\sim
2-3$), that indicates to the transition from the fragmentation
mechanism to the recombination one.The recombination mechanism
provides justification for applying the statistical model to
describe ratios of hadron yields (the ratio $Baryon/Meson\approx
1$) \cite{HWA}. The collective flow of quarks may be explained by
the recombination mechanism, too. The rapid local thermalization
may be a consequence of this formation of secondary hadrons
\cite{HWA}.

In this way we try to compare two kinds of  processes which have
different values of hadronization parameters. The first one is
$e^+e^-$ - annihilation. It is usually supposed that fragmentation
dominates in it and newly formed hadrons fragment with a high
moment of parton into the surrounding vacuum  (such objects can
also appear from the hot surface in peripheral events in nucleus
and hadron collisions) \cite{HWA}.

The nuclear modification factor $R_{CP}$  and elliptic flow $v_2$
in Au-Au collisions at RHIC have revealed an apparent quark-number
dependence in the $p_T$ region from 1.5 to 5 GeV/c. Moreover, the
baryon production increases more rapidly with centrality than the
meson production. These observations confirm the picture of hadron
formation by  quark recombination \cite{HWA} and point out that
the hadronization processes in high energy nucleus interactions
are modified  to the comparison of $e^++e^-$ and partly  $p+p$
collisions.

The GDM with a branch gives growth of the part of the evaporated
gluons to 0.85-0.98 and a small rise of gluon branch number at
higher energies. Besides we have got data about emergence of hard
constituent in MP \cite{GIO}. In GDM it can be explained not only
by not only evaporation of a single gluon sources but also of
groups with several gluons (formed by branch). A simple MD scheme
of this  superposition will be analyzed below.

Let us compare MD (\ref{38}) with the descriptions of experimental
data obtained by various approaches. We bring two of them. A
fortunate expression for KNO function was obtained by a group from
IHEP \cite{SEM} who combined the elastic and inelastic processes.
We can see (Fig. 9) good agreement with data \cite{PrD} at 800
GeV/c both of MD in MGD (solid line) and KNO-function (dot line).

A wide research of MD in pp-interactions was fulfilled by L.Van
Hove, A.Giovannini and R.Ugoccioni \cite{GIO}. They proposed a
two-step mechanisms of MP. The independent (Poisson) production of
groups of ancestor particles (named "clan ancestors") were
supplemented by their decay, according to a hadron shower process
(the logarithmic MD within each clan). Such convolution of two
mechanisms gives a negative binomial distribution (NBD) for
hadrons
\begin{equation}
\label{45} P_n(s)=\frac{k_h(k_h+1)\dots(k_h+
n-1)}{n!}\left(\frac{\overline n(s)} {\overline
n(s)+k_h}\right)^{n}\left( \frac{k_h}{k_h+\overline
n(s)}\right)^{k_h},
\end{equation}
where $k_h$ - the NBD parameter and $\overline n(s)$ - the mean
multiplicity of hadrons. The comparison of NBD (dot line) and our
MD in GDM (solid line) with data at 800 Gev/c is given in Fig. 10.
A.~Giovannini emphasizes that the nature of this clan is gluon
bremsstrahlung \cite{GIO}. Our investigations by GDM allows to
give a concrete gluon content. Binomial distributions (BD)
describe the hadronization stage. The clan model of \cite{GIO}
uses the logarithmic distribution of secondaries in a single clan.
Both of MD have the similar behavior.

At the top energy (especially at 900 GeV) the shoulder structure
appears in $P_n$ \cite{UA5}. The comparison of data with one NBD
does not describe data well. But the weighted superposition of two
NBD gives a good description of the shoulder structure $P_n(s)$
\cite{GIO}. At 14 TeV A.Giovannini expects the weighted
superposition of the three classes of events.

We can modify our GDM considering that the gluon fission may be
realized at higher energies. The independent evaporation of gluons
sources of hadrons may be realized by single gluons and also
groups from two and more fission gluons. Following A.Giovannini we
name such groups of gluons - clans. Their independent emergence
and following hadronization content of GDM. MD in GDM with two
kinds of clans are:
$$ P_n(s)=\alpha _1\sum\limits_{m_1=0}^{Mg_1}\frac{e^{-\overline
m_1} \overline m_1^{m_1}}{m_1!} C^{n-2}_{m_1\cdot
N}\left(\frac{\overline n^h}
{N}\right)^{n-2}\left(1-\frac{\overline n^h} {N}\right)^{m_1\cdot
N-(n-2)}+
$$
\begin{equation}
\label{48} + \alpha
_2\sum\limits_{m_2=0}^{Mg_2}\frac{e^{-\overline m_2} \overline
m_2^{m_2}}{m_2!} C^{n-2}_{2\cdot m_2\cdot N}\left(\frac{\overline
n^h} {N}\right)^{n-2}\left(1-\frac{\overline n^h}
{N}\right)^{2\cdot m_2\cdot N-(n-2)},
\end{equation}
where $\alpha _1$ and $\alpha _2$ are the contribution single and
double gluon clans ($\alpha _1 + \alpha _2 =1$). The comparison
(\ref{48}) with experimental data for proton interactions at
$\sqrt s= 62.2$ GeV \cite{ISR} is given in Fig. 8. We have
obtained the following values of parameters: $N=7.06\pm 3.48$,
$\overline m_1 =3.59 \pm 0.03$, $\overline m_2=1.15\pm 0.25$,
$\overline n_h=3.23\pm 0.14$, $Mg_1=8$, $Mg_2=4$, $\alpha
_1/\alpha _2 \sim 1.8$ at $\chi ^2/$ndf=$9.12/13$. The mean
multiplicities of the two kinds of clans are similar.

The specific feature of our GDM approach is the dominance of a lot
of active gluons in MP. We can expect the emergence of them in
nucleus collisions (experiments at RHIC) and the formation of a
new kind of matter (quark-gluon plasma)at high energy. We consider
that our gluon system can be a candidate for this. So the mean
multiplicity of active gluons approached 10 at RHIC. For Au+Au
central collisions their number may be equal to $ 200\cdot
\overline m \approx 2000$ before the branch. This gluon medium
facilitates the quenching.
\begin{figure}
\begin{minipage}[b]{.3\linewidth}
\centering
\includegraphics[width=\linewidth, height=2in, angle=0]{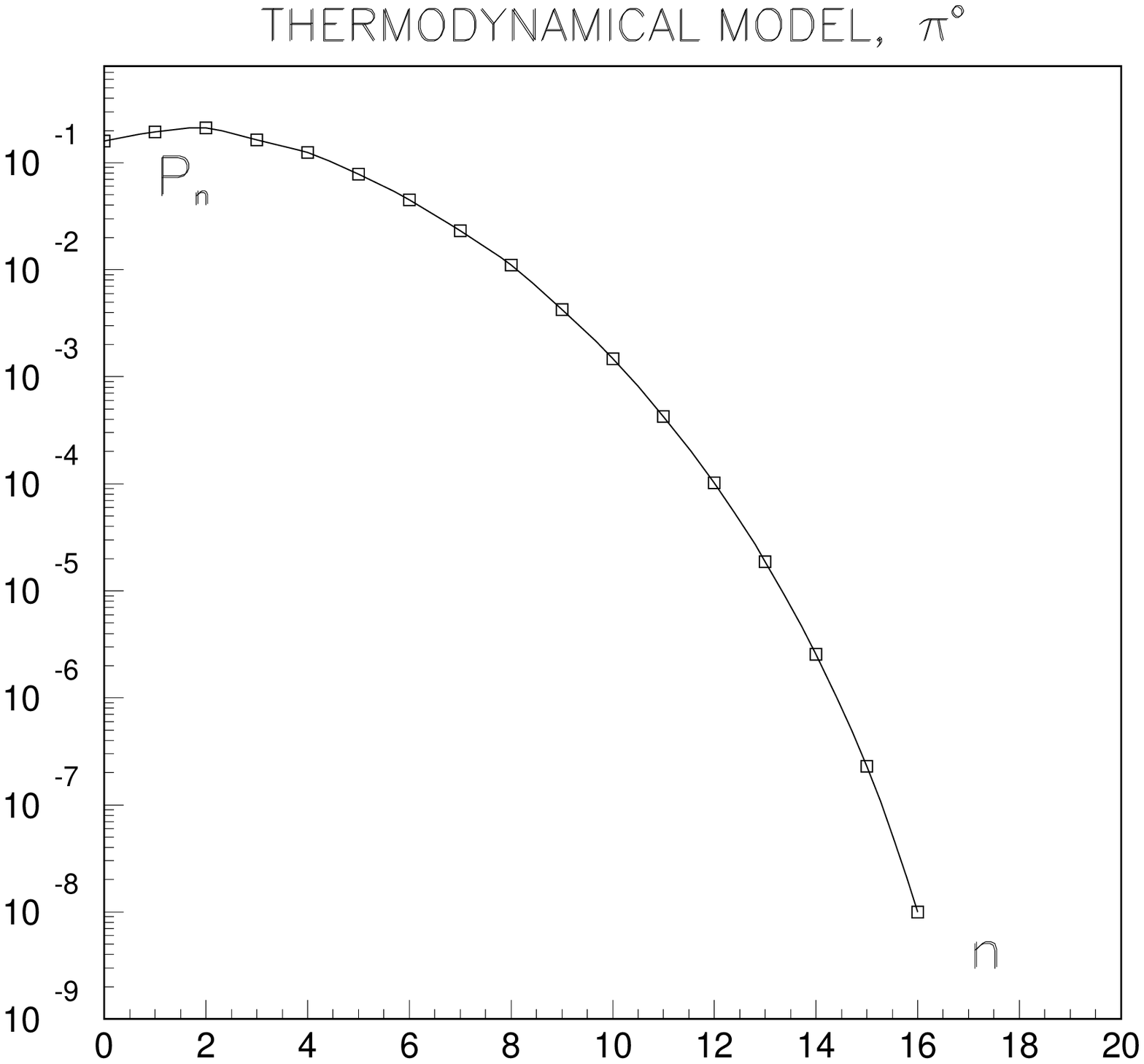}
\caption{MD for $\pi ^0$.} \label{25kdfig}
\end{minipage}\hfill
\begin{minipage}[b]{.3\linewidth}
\centering
\includegraphics[width=\linewidth, height=2in, angle=0]{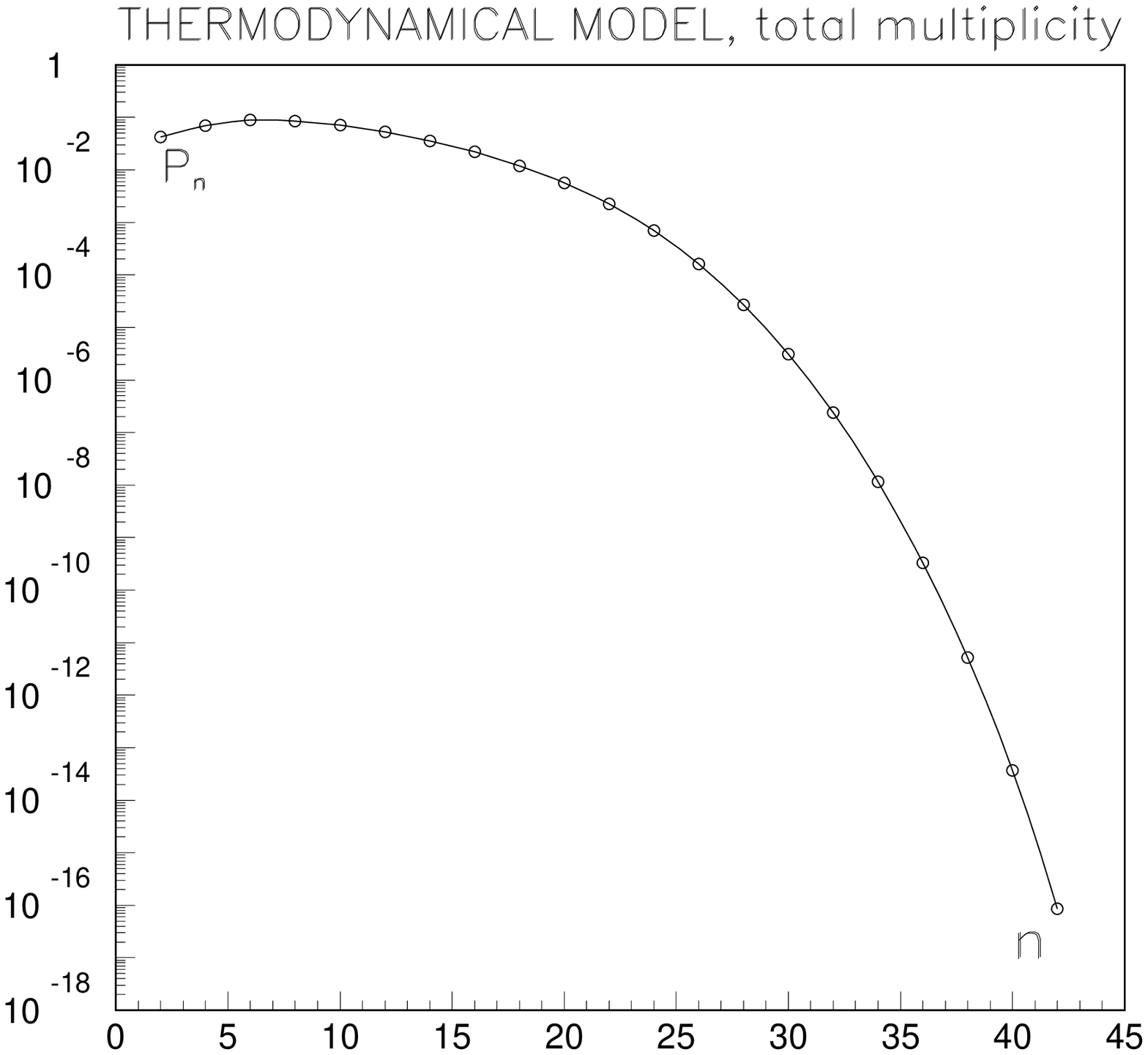}
\caption{MD for $n_{tot}$.} \label{26kdfig}
\end{minipage}\hfill
\begin{minipage}[b]{.3\linewidth}
\centering
\includegraphics[width=\linewidth, height=2in, angle=0]{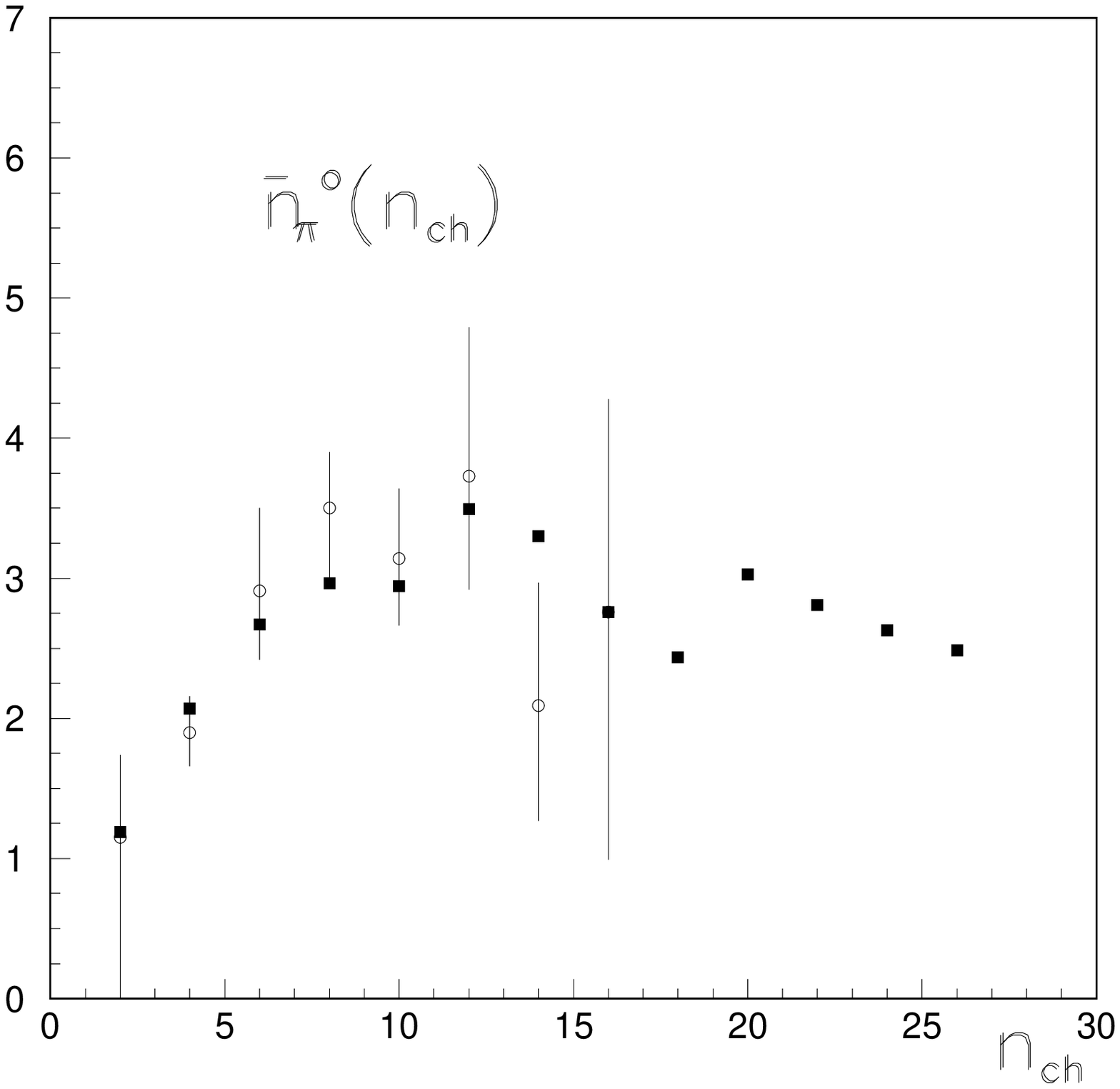}
\caption{ $\overline n_\pi^o(n_{ch})$.} \label{27kdfig}
\end{minipage}
\end{figure}
\section{MD in $p\bar p$-annihilation}
In the midst of interesting and enough inextricable hadron
interactions the $p\bar p$ annihilation shows up especially
\cite{Rush}. Experimental data at tens GeV/c \cite{Rush} point out
on some maxima in differences between $p\bar p$ and $pp$ inelastic
topological cross sections what may witness about the contribution
of different mechanisms of MP
\begin{equation}
\label{50} \Delta \sigma _n(p\overline p -pp)= \sigma _n(p
\overline p) - \sigma _n (pp).
\end{equation}
The important information about the MP mechanism may be picked out
from the MD moment analysis of charged particles. The second
correlative moment for negative particles $f_2^{--}$ are available
to study MP
\begin{equation}
\label{49} f_2^{--}=\overline {n_-(n_--1)}-{\overline{n}_-}^2.
\end{equation}
The negative value of second correlative moments is characteristic
for a more narrow MD than Poisson, and they indicate the
predominance of the hadronization stage in MP. According to MGD,
active gluons are a basic source of secondary hadrons.

At the initial stage of annihilation three valent $q\overline
q$-pairs ($uud$ and $\overline u \overline u \overline d$) are.
They can turn to the "leading" mesons which consist from: a)
valent quarks or b) valent and vacuum quarks \cite{KUR}. In the
case a) only three "leading" neutral pions (the "0" topology) or
two charged and one neutral "leading" mesons ("2" - topology) may
form. In b) case the "4"- and "6"- topology is realized for
"leading" mesons. We suggest that the formation neutron and
antineutron (exchange) can be realized.

A simple scheme of MP for annihilation may give the negative
second correlative moments in GDM. We suggest that the active
gluon emergence together with the formation of intermediate
topology occurs. The GF for a single active gluon
$Q_1(z)=[1+\overline n/N(z-1)]^N$ gives \cite{MGD}
\begin{equation}
\label{52} f_2=Q_1^{''}(z)|_{z=1} -[Q_1(z)|_{z=1}]^2 = -(\overline
n^h)^2/N < 0.
\end{equation}
Reciprocally for m gluons GF and $f_2$ will be
\begin{equation}
\label{53} Q_m(z)=[1+\overline n^h/N(z-1)]^{mN}, \quad
f_2=-m(\overline n^h)^2/N.
\end{equation}
We consider that $m$ grows while increasing the energy of the
colliding particles, and $f_2$ will decrease  almost linearly from
$m$. Such behavior qualitatively agrees with experimental data
\cite{Rush}. If we take concrete MD $P_m^G$ for gluons, then GF
for secondary hadrons and $f_2$ are
\begin{equation}
\label{54} Q(z)=\sum\limits_m P_m^G [1+\overline n^h/N(z-1)]^{mN}
\end{equation}
\begin{equation}
\label{55} f_2=[f_2^G +1-1/N]\cdot \overline m \cdot(\overline
n^h)^2,\quad f_2^G=\overline {m(m-1)}- \overline m^2,
\end{equation}
where $f_2^G$ - the second correlative moment for gluons. In this
scheme $f_2$ may be negative or positive. We consider that the
negative value $f_2$ in the large energy region in comparison with
$p+p$ interactions may be related with the destruction of the
initial system on three or more shares and the number of active
gluons related with a "leading" pion will be less than in the case
of a leading proton in pp-collisions at the same energy. Herewith
the total number of such gluons at annihilation may be bigger,
their manifestation happens independently but the number of them
per one pion grows slowly. The explanation of the negative $f_2$
was given R.Lednicky \cite{Led} at the assumption of the
independent MP of charged particles. The second correlative moment
has a zero value only in the small energy domain. And so we should
restrict the region to apply this explanation.

According to GDM for $p\overline p$ annihilation and taking into
account three intermediate charged topology and active gluons, GF
$Q(z)$ for final MD may be written as the convolution gluon and
hadron components:
$$Q(z)=c_0\sum\limits_m
P_m^G [1+\frac{\overline n^h}{N}(z-1)]^{mN}+c_2\sum\limits_m z^2
P_m^G [1+\frac{\overline n^h}{N}(z-1)]^{mN} +c_4\sum\limits_m z^4
P_m^G [1+\frac{\overline n^h}{N}(z-1)]^{mN}.
$$
The parameters of $c_0$, $c_2$  and $c_4$ are determined as the
part of intermediate topology ("0", "2" or "4") to the
annihilation cross section ($c_0 + c_2 + c_4 =1).$ For the
simplicity we are limited by Poisson distribution with the finite
number of gluons for $P_m^G$.

The comparison of the  experimental data (Fig. 12) gives the
following values of parameters: $\overline m=3.36 \pm 0.18$,
$N=4.01\pm 0.61$, $\overline n^h$=1.74 $\pm $0.26, the ratio $c_0$
: $c_2$ : $c_4$ = 15 : 40 : 0.05 at $\chi ^2/ndf=5.77/4$ and the
maximum possible number of gluons $M=4$ at "4"-topology. The sum
begins from $m=1$ (inelastic events), at $n \geq 2$ - from $m=0$
and finishes up $m<M$ at small multiplicities ($n\leq 4$). We
should to emphasize very complicated events $n_{ch}=0$ and $2$.
This research of $p\overline p$ annihilation requares to be
continued. We will develop MGD to describe MD at energies 200,
500, 900 GeV \cite{ALN} and higher.

\begin{figure}
\begin{minipage}[b]{.3\linewidth}
\centering
\includegraphics[width=\linewidth, height=2in, angle=0]{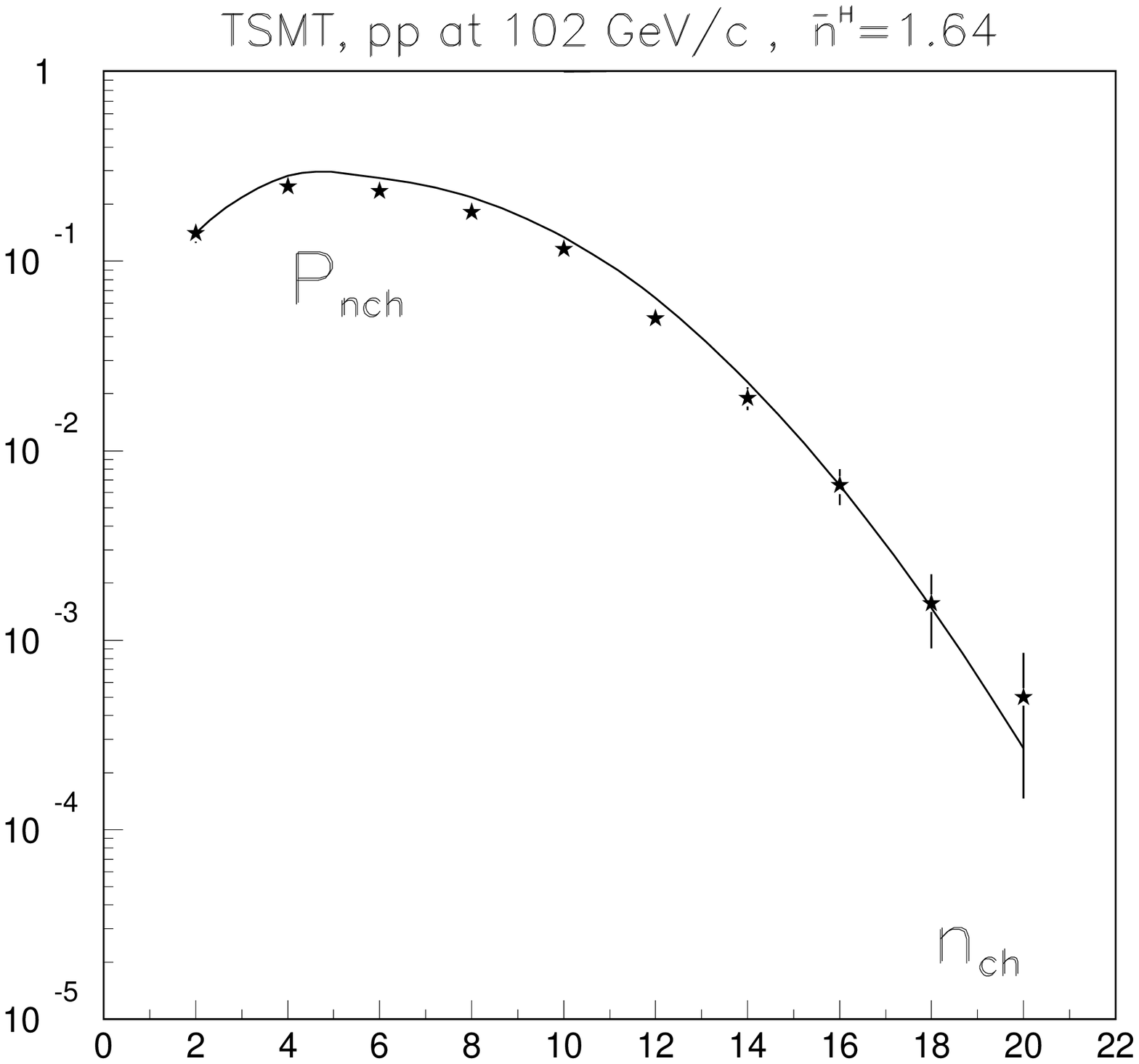}
\caption{MD at 102 GeV/c.} \label{31dfig}
\end{minipage}\hfill
\begin{minipage}[b]{.3\linewidth}
\centering
\includegraphics[width=\linewidth, height=2in, angle=0]{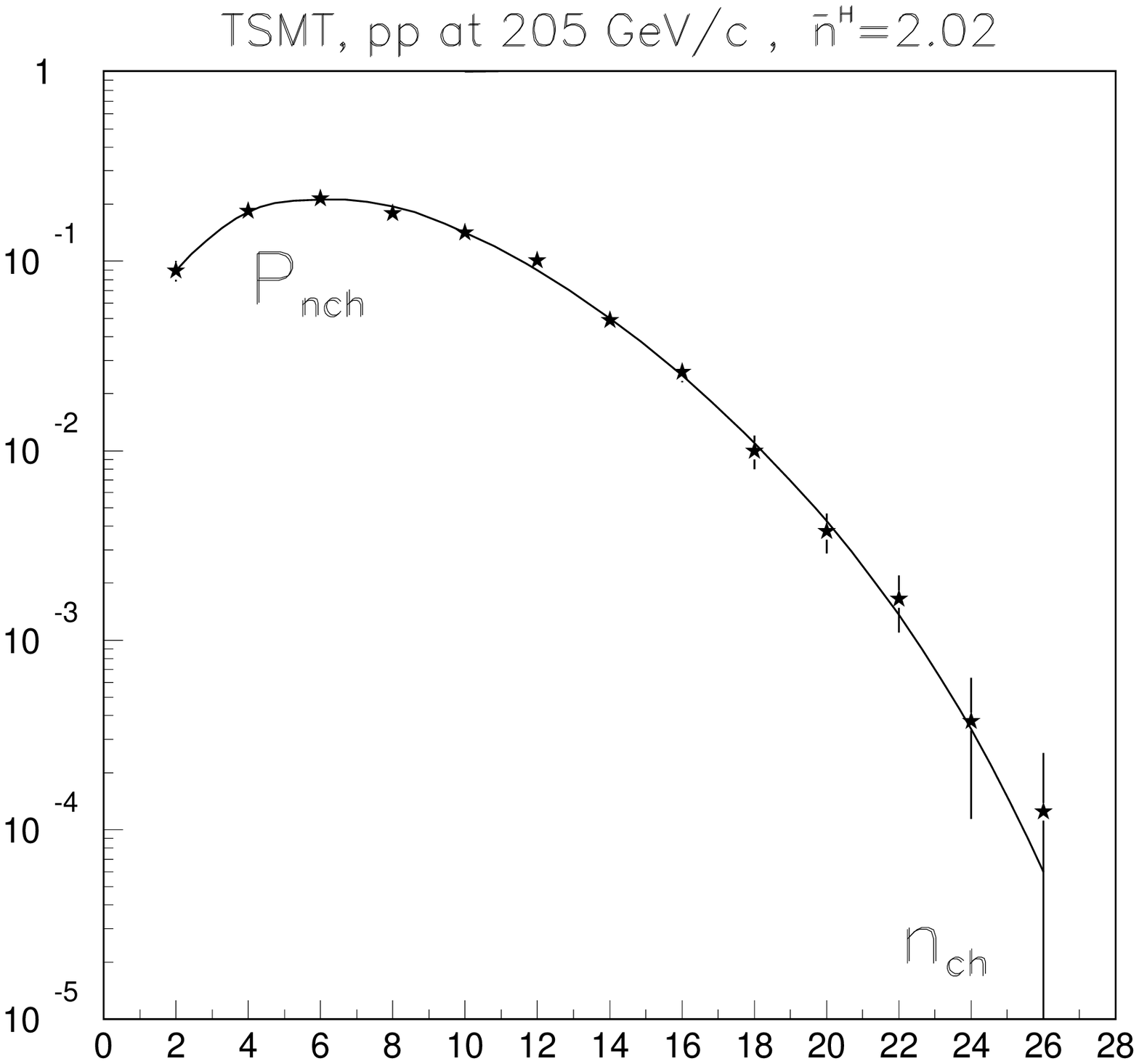}
\caption{MD at 205 GeV/c.} \label{32dfig}
\end{minipage}\hfill
\begin{minipage}[b]{.3\linewidth}
\centering
\includegraphics[width=\linewidth, height=2in, angle=0]{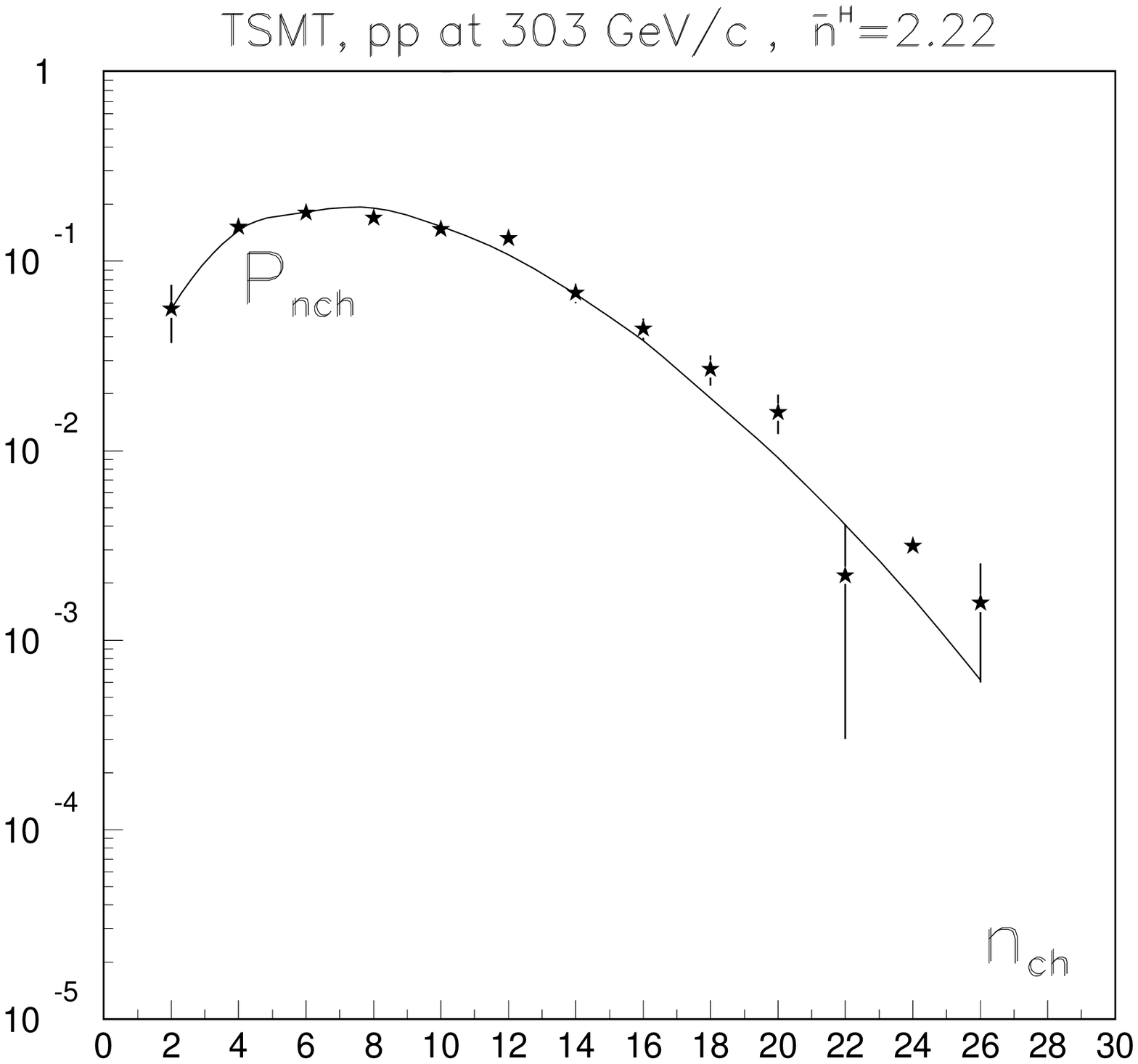}
\caption{MD at 303 GeV/c.} \label{33dfig}
\end{minipage}
\end{figure}

\begin{figure}
\begin{minipage}[b]{.3\linewidth}
\centering
\includegraphics[width=\linewidth, height=2in, angle=0]{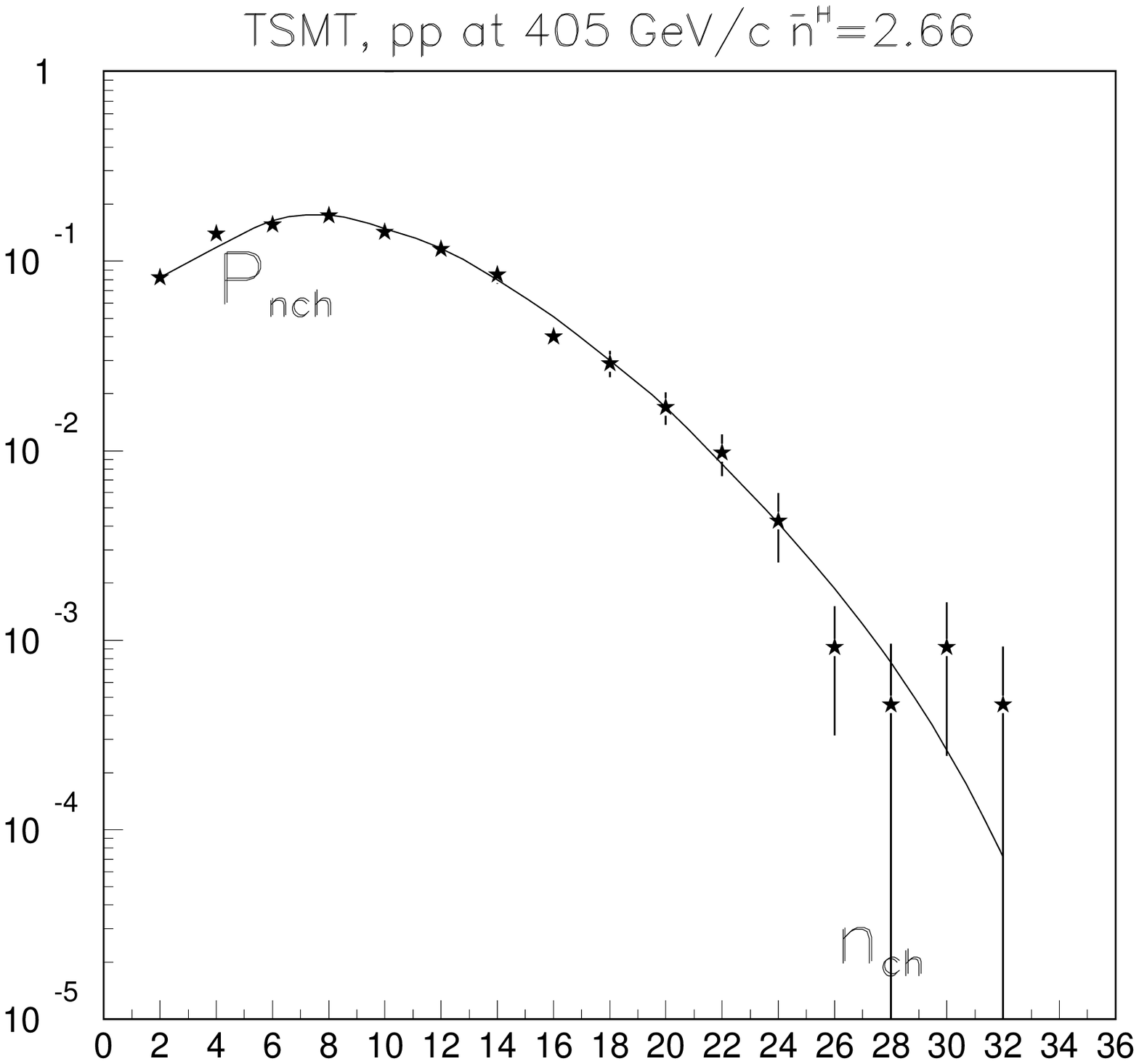}
\caption{MD at 405 GeV/c.} \label{34dfig}
\end{minipage}\hfill
\begin{minipage}[b]{.3\linewidth}
\centering
\includegraphics[width=\linewidth, height=2in, angle=0]{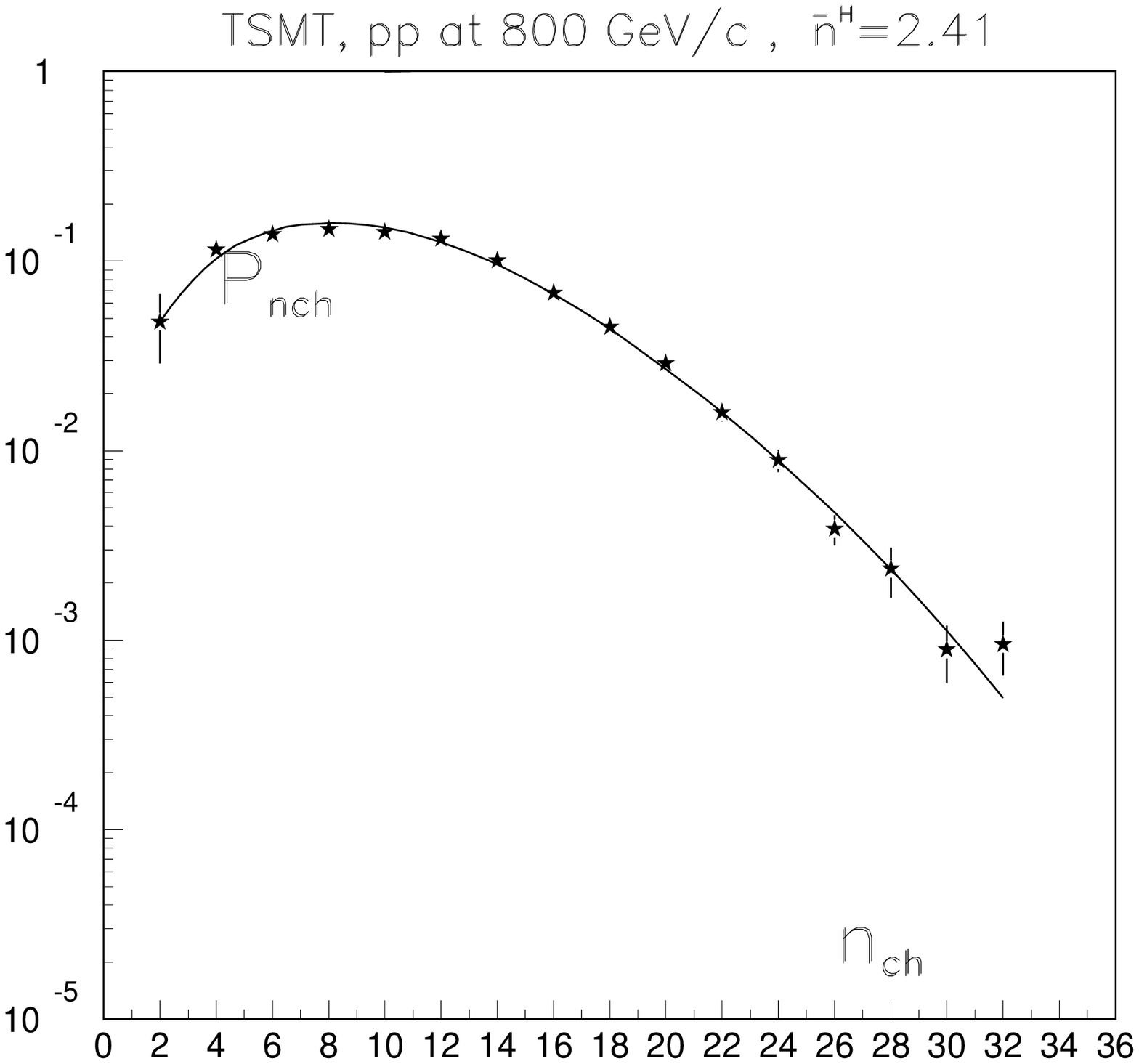}
\caption{MD at 800 GeV/c.} \label{35dfig}
\end{minipage}\hfill
\begin{minipage}[b]{.3\linewidth}
\centering
\includegraphics[width=\linewidth, height=2in, angle=0]{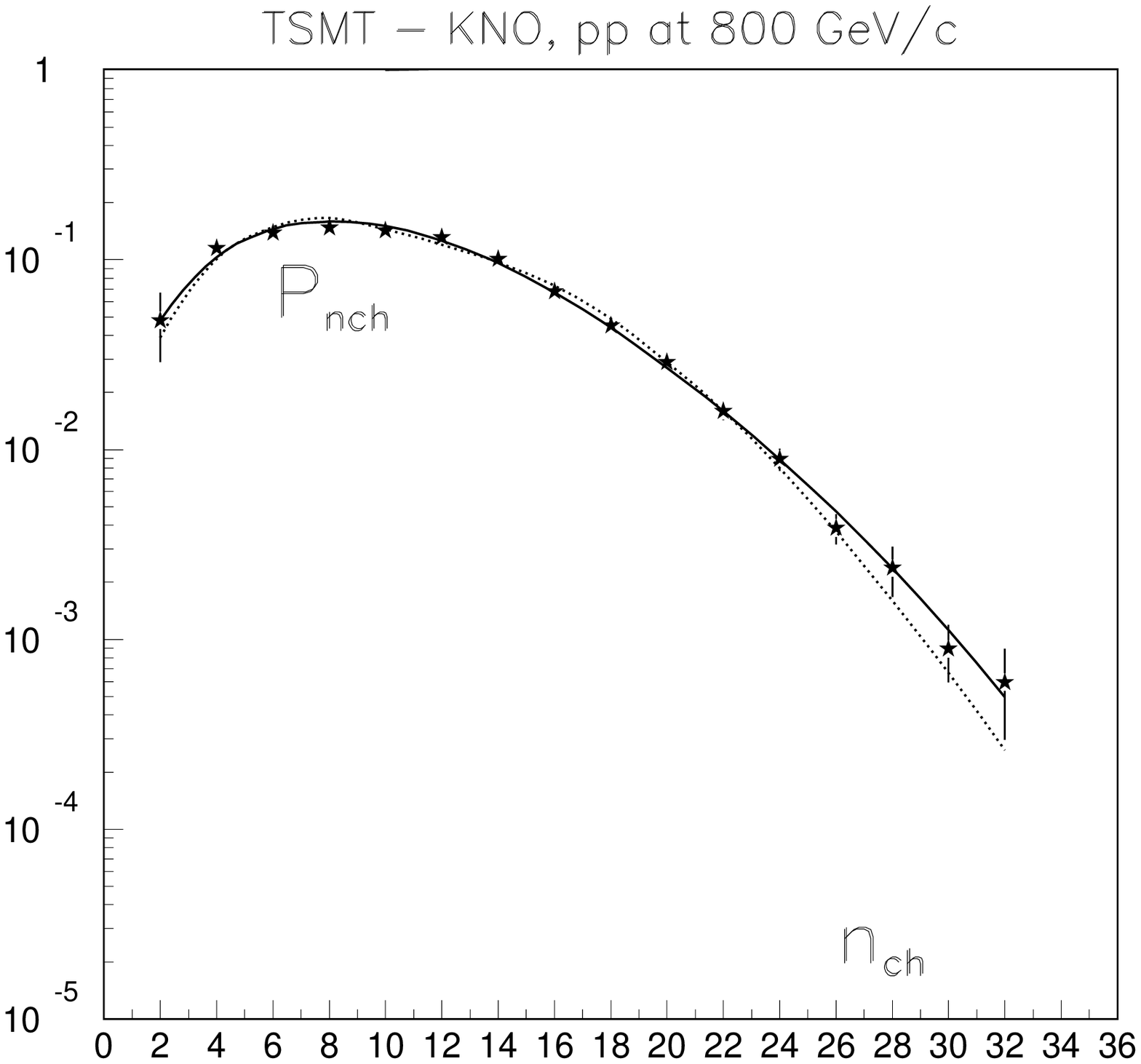}
\caption{MD in GDM and KNO-function \cite{SEM}.} \label{36dfig}
\end{minipage}
\end{figure}
\section{Soft photons}
The production of photons in particle collisions at high energies
was studied in many experiments \cite{CHL}. In project
"Thermalization" it is planned to investigate low energetic
photons with $p_t \leq 0.1GeV/c$ and $x \leq 0.01$ \cite{THE}.
Usually these photons are named soft photons (SP). Experiments
shown that measured cross sections of SP are several times larger
than the expected ones from QED inner bremsstrahlung.
Phenomenological models were proposed to explain the SP excess:
the glob model of Lichard and Van Hove and the modified soft
annihilation model of Lichard and Thomson \cite{LVH} .

We consider that at a certain moment QGS or excited new hadrons
may set in an almost equilibrium state during a short period or
finite time. That is why, to describe massless photons, we will
try to use the black body emission spectrum \cite{HER}. From
experimental data \cite{THE} the inelastic cross section is equal
to approximately $40 mb$, the cross section of SP formation is
about $4 mb$, and since $ \sigma _{\gamma} \simeq n_{\gamma
}(T)\cdot \sigma _{in}$, then the number of SP will be equal to
$n_{\gamma }\approx 0.1.$ For convenience, we may use the
well-known density of MVB at $T_r=2.275 K$ and get the number of
photons by means of MVB $n_{\gamma }(T)=n_{\gamma }(T_r)\cdot
\left ( \frac {T}{T_r}\right )^3.$ The density of SP in the region
$1 fm^3$ will be equal to
$$
\rho (T)=n_{\gamma }(T)/V=4.112\cdot 10^8\cdot 10^{-6}\cdot
10^{-39}\cdot \left ( \frac {T}{T_r}\right)^3 fm^{-3}.
$$
The estimates of temperature are implemented by transfer moment:
$T=p\approx p_T\sqrt 2$ ($1MeV=1.16\cdot 10^{10}K$). If $T(p_T)$
is known, using $n_{\gamma }$ we can estimate the linear size of
radiation system ($V\simeq L^3$).
 Dependencies of the linear size of system (L) from the
SP moment ($p_T$) are given in Table~ 2.
\begin{center}
Table 2. The size of system L (fm) versus $p_T$ (MeV/c) of SP.
\end{center}
\renewcommand{\tablename}{Table}
\begin{center}
\begin{tabular}{|c|c|c|c|c|c|c|}
\hline \hline $ \quad p_T \quad $ &$\quad  10 \quad $ &$
\quad 15 \quad  $&$ \quad  25 \quad  $&$ \quad 30 \quad$ &$ \quad 40 \quad$&$ \quad 50 \quad$ \\[1ex]
\hline \hline
$L$ &$11$&$6.9$&$4.1$&$3.5$&$2.6$&$2.0$\\[1ex]
\hline \hline
\end{tabular}
\end{center}

It is well-known that the temperature of second hadrons is higher
than the temperature of SP. We presume that objects with soft
gluon content may not transform into hadrons but turn into SP. The
amount of such soft gluons is estimated by $N_g$ in TSMB.

\begin{figure}
\begin{minipage}[b]{.3\linewidth}
\centering
\includegraphics[width=\linewidth, height=2in, angle=0]{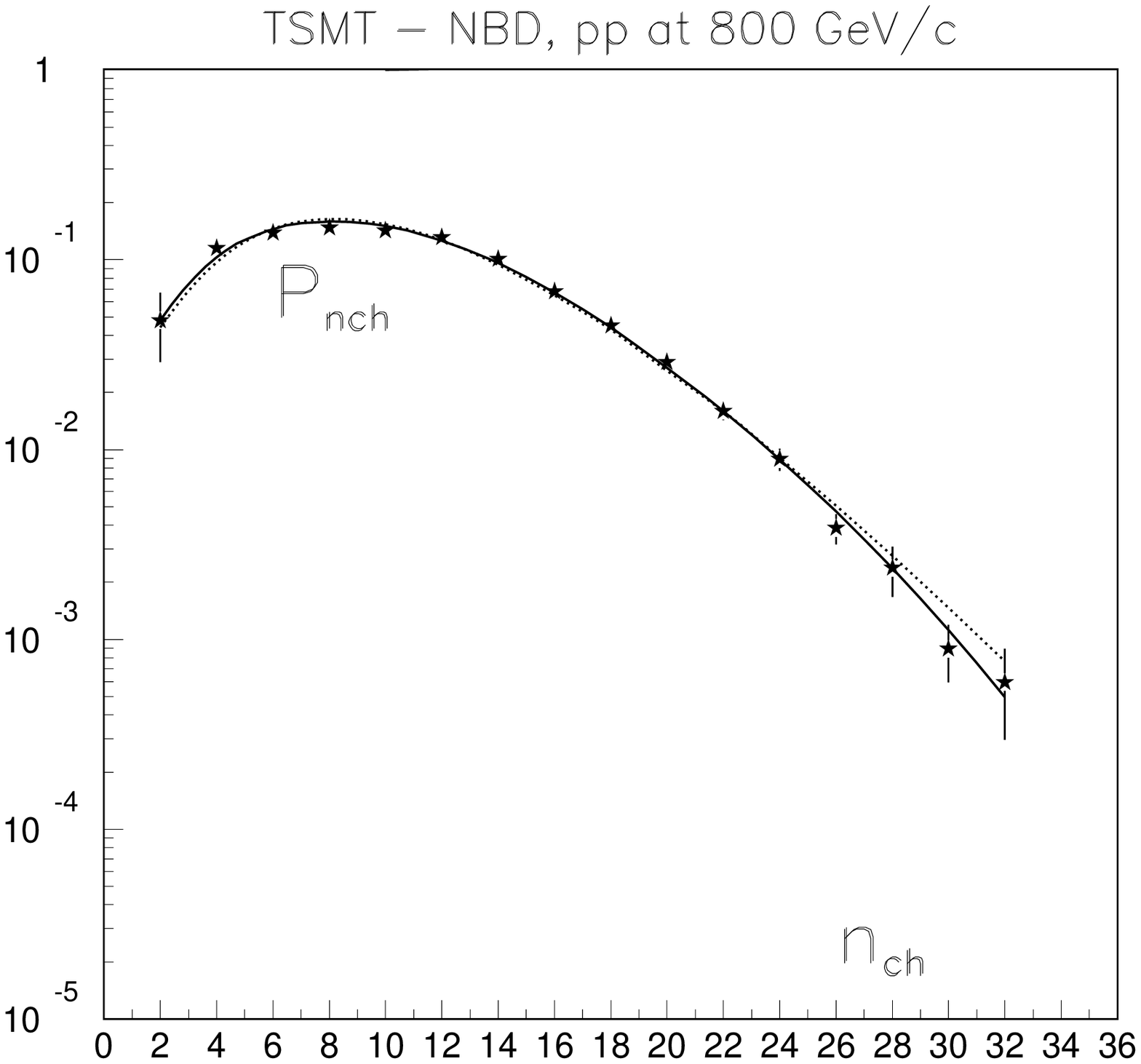}
\caption{MD in TSTM and NBD.} \label{37dfig}
\end{minipage}\hfill
\begin{minipage}[b]{.3\linewidth}
\centering
\includegraphics[width=\linewidth, height=2in, angle=0]{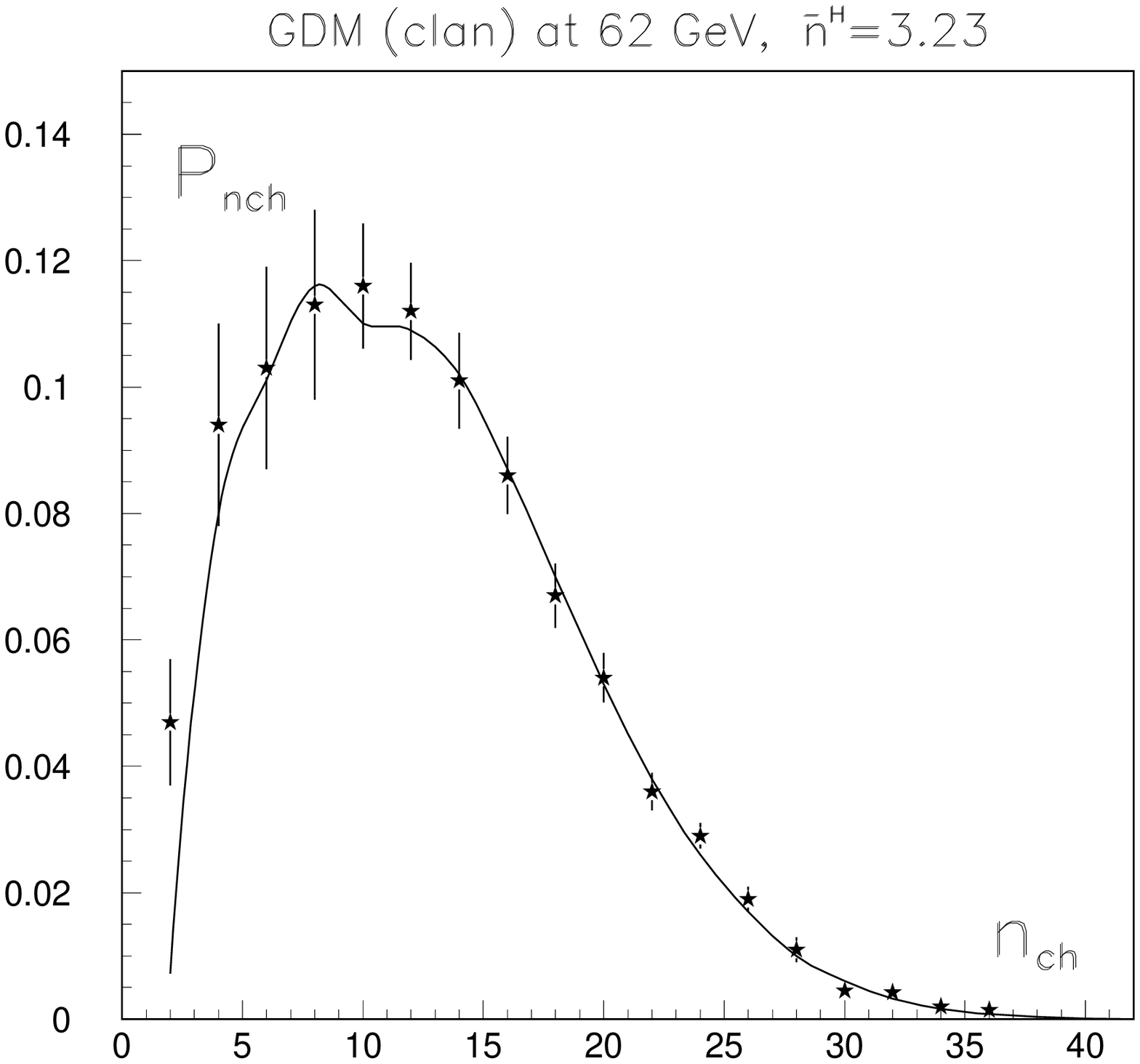}
\caption{MD in GDM (clan).} \label{38dfig}
\end{minipage}\hfill
\begin{minipage}[b]{.3\linewidth}
\centering
\includegraphics[width=\linewidth, height=2in, angle=0]{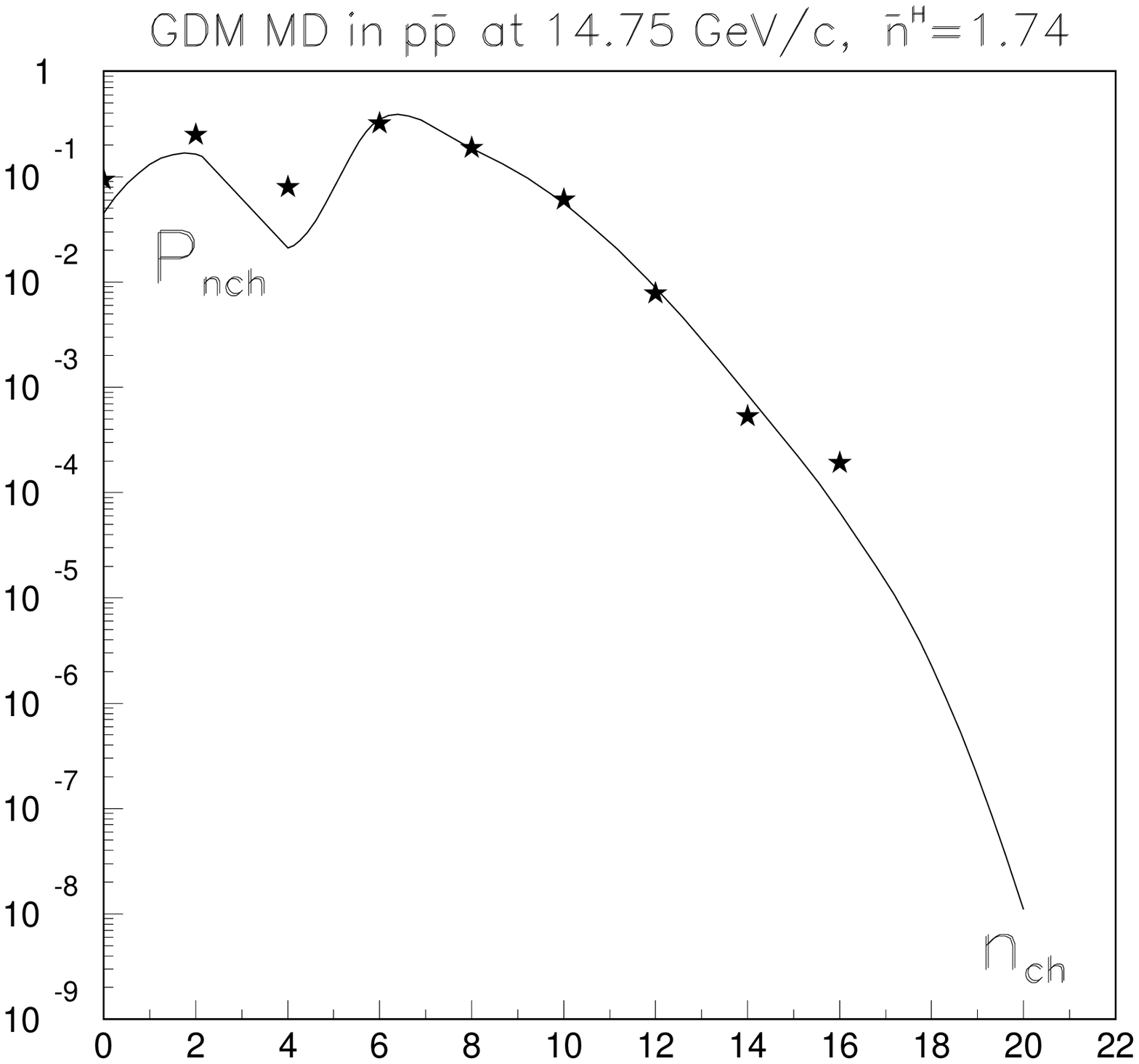}
\caption{GDM MD in $p\overline p$ at 14.75 GeV/c.} \label{39dfig}
\end{minipage}
\end{figure}
\section{Conclusion}
In our research we have undertaken an attempt to give MP
description in different processes by means of a unified approach
based on quark-gluon picture using the phenomenological
hadronization model. The implemented model investigation allows us
to understand deeper the picture of MP at various stages. We have
obtained qualitative and quantitative agreements of our schemes
with experimental data in $e^+e^-$, $p\overline p$ annihilation
and $pp$ and nucleus collisions in a very wide energy domain.

The authors appreciate for the support of physicists from JINR,
GSTU who encouraged our investigations. {\it Project
"Thermalization" is partially supported by RFBR grant
03-02-16869}.

\end{document}